\documentclass[amsmath,amssymb,twocolumn]{revtex4}

\usepackage{graphicx}% Include figure files
\usepackage{dcolumn}% Align table columns on decimal point
\usepackage{bm}% bold math

\begin{document}

\title{Outward Accessibility in Urban Street Networks:
Characterization and Improvements}
\author{Bruno Augusto Nassif Traven\c{c}olo}
\author{Luciano da Fontoura Costa}
\affiliation{%
Institute of Physics at S\~ao Carlos, University of S\~ao Paulo,\\
PO Box 369, S\~ao Carlos, S\~ao Paulo, 13560-970 Brazil }%
\date{\today}

\begin{abstract}
The dynamics of transportation through towns and cities is strongly
affected by the topology of the connections and routes.  The current
work describes an approach combining complex networks and
self-avoiding random walk dynamics in order to quantify in objective
and accurate manner, along a range of spatial scales, the
accessibility of places in towns and cities.  The transition
probabilities are estimated for several lengths of the walks and
used to calculate the outward accessibility of each node. The
potential of the methodology is illustrated with respect to the
characterization and improvements of the accessibility of the town
of S\~ao Carlos.
\end{abstract}

\maketitle

The intrinsic relationship between structure and dynamics seems to
scaffold many dynamical processes in nature, from the flight of birds
to the binding of proteins.  Because of its inherent ability to
represent and model the most diverse types of discrete structures,
complex networks have received growing attention. Having initially
focused attention on the characterization and modeling of the topology
of interconnectivities (e.g.~\cite{Albert2002,Newman2003,Costa2007a}),
complex network research progressed steadily to encompass the
relationship between structure and dynamics in the most diverse
systems (e.g.~\cite{Newman2003,Costa2007b}). Though the connectivity
does not completely define dynamics, it strongly affects it.  This has
become clear through investigations of relationships between structure
and several types of dynamics, including diffusion
(e.g.~\cite{Costa2007}), synchronization (e.g.~\cite{Boccaletti2006})
and neuronal networks~\cite{Costa2003,Costa2007c}. Particularly, when
the dynamics is modeled in terms of random walks
(e.g.~\cite{Costa2007, Gardenes2007}), the displacements of the
respective moving agents are strongly influenced by several
topological factors such as the number of connections at each node and
the shortest path lengths between nodes.  This type of stochastic
dynamics presents an intrinsic potential for modeling the displacement
of people within towns or cities.  However, because traditional linear
random walks allow a moving agent to visit edges and nodes more than
once, implying null average displacement in the long term, it becomes
important to consider more purposive types of displacements.
Self-avoiding walks (e.g.~\cite{Herrero2003, Herrero2005, Yang2005,
Kinouchi2002}) represent a natural simplified choice for modeling
urban displacements, implying the agents to move away from their
initial position in a more effective way while not repeating edges or
nodes.

\begin{figure*}[ht]
\includegraphics{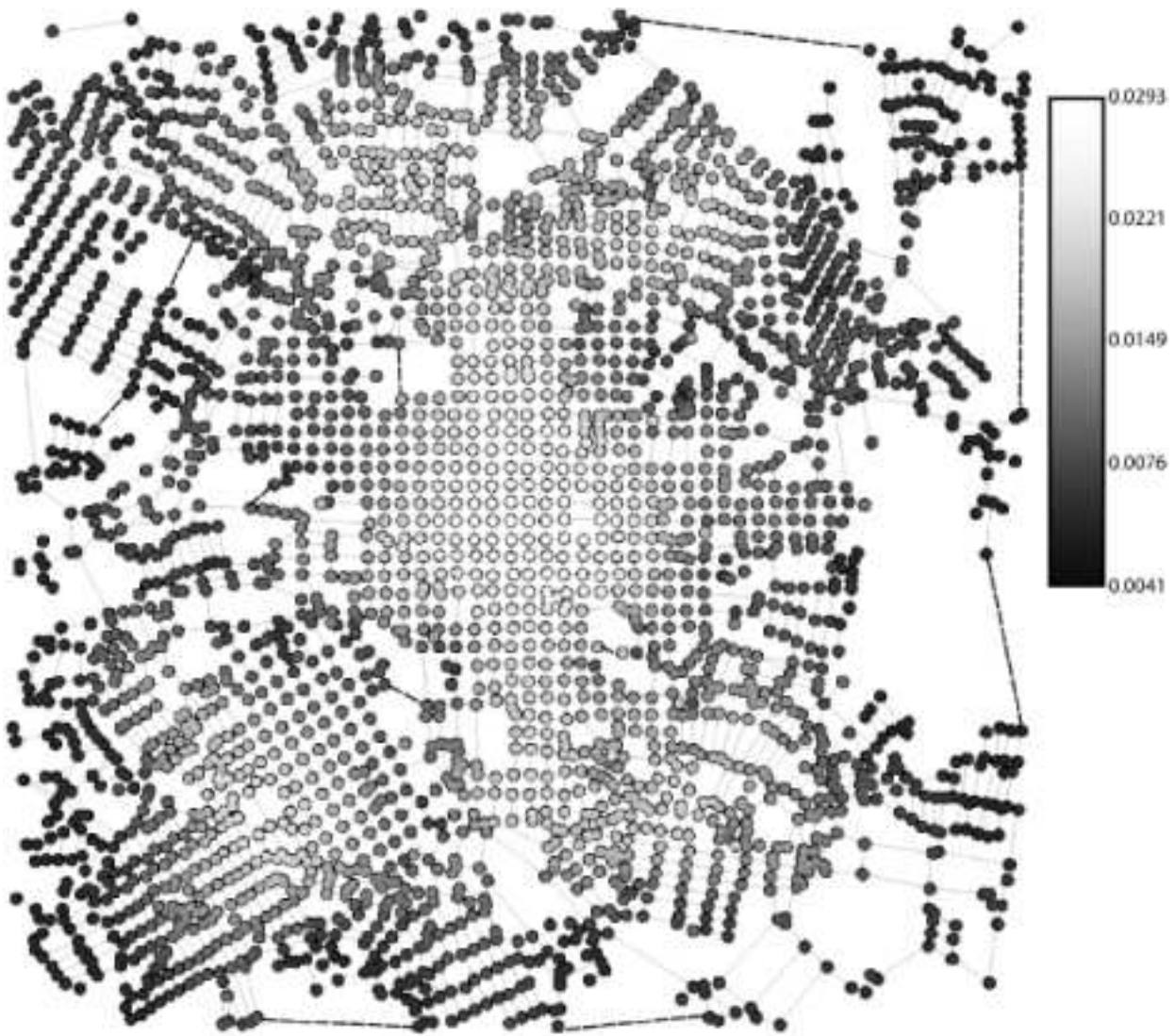}
\caption{\label{fig1} Network of urban streets of S\~ao Carlos,
Brazil. The gray level of the nodes indicates their respective
averaged outward accessibility accordingly to the legend at the
right-hand side. The dashed lines represent the hypothetical
additional edges aimed at improving the accessibility.}
\end{figure*}

Complex networks have been used to characterize important
topological, dynamical and spatial properties of cities (e.g.,
~\cite{Rosvall2005, Crucitti2006, Cardillo2006, Lammer2006,
Porta2006}). One important practical application of the
structure-dynamics relationship concerns the characterization,
modeling and planning of urban displacements. In a previous study,
Rosvall et al.~\cite{Rosvall2005} considered shortest path lengths
in order to quantify and compare the information needed to locate
specific addresses in different cities. The current work applies the
recently introduced concept of
\emph{accessibility}~\cite{Costa2008b} in order to quantify in an
objective and comprehensive way the outward accessibilities of each
node of a town (i.e. intersection or beginning of routes). The
methodology is illustrated with respect to the specific application
to the Brazilian town of S\~ao Carlos. Image processing and analysis
methods were used to transform the plan of the town into a
respective geographical planar network, where the nodes represent
the crossings and the beginning of routes, while the edges
correspond to the streets. Figure~\ref{fig1} shows the network
derived from the central part of the town.  Then, by simulating a
series of self-avoiding walks starting at all nodes, transition
probabilities from one node to another were estimated with respect
to varying lengths of the walks. The \emph{outward accessibility} of
each node --- expressing the diversity of routes between those
points as well as the potential of the moving agent to visit a set
of nodes in the shortest time --- is quantified in terms of the
entropy of the obtained transition probabilities, so that values
close to $1$ indicate maximum outward accessibility.  One important
property of the outward accessibility measurement is that
self-avoiding walks initiating from nodes characterized by high
accessibility for a given path length tend to visit all reachable
nodes at that length in the shortest period of time.  In addition,
the outward accessibility intrinsically considers the number of
alternative routes from the initial node to the reachable nodes.
Nodes with high outward accessibility therefore have more balanced
number of routes leading to the reachable nodes~\cite{Costa2008a,
Costa2008b}.

We start by presenting the basic concepts about network
representation. An unweighted and undirected complex network can be
represented by a matrix $K$, called \emph{adjacency matrix}. The
dimension of this matrix for a network with $N$ nodes and $E$ edges
is $NxN$. If the nodes $i$ and $j$ are connected through an edge,
the elements $K(i,j)$ and $K(j,i)$ of the adjacency matrix are set
to $1$; otherwise $K(i,j) = K(j,i) = 0$. Two nodes of the network
are said to be adjacent if they share one edge. Two edges of the
network are said to be adjacent if one extremity of each edge share
the same node. The \emph{degree} of a node $i$ is the number of its
immediate neighbors. A \emph{walk} over the network is composed by a
sequence of adjacent nodes, starting from an initial node and
proceeding through successive steps $h$. A \emph{self-avoiding walk}
is a walk where the nodes and edges do not appear more than once.

After a real-world structure has been properly represented as a
complex network, a diversity of measures can be obtained, ranging from
simple features such as the node degree and clustering coefficient, to
more sophisticated such as shortest path lengths and betweeness
centrality. These measures have allowed the comprehensive description
and characterization of several complex
systems~\cite{Costa2007a,Costa2007b}.

The \emph{transition probability} that a moving agent node reaches a
node $j$ after departing from a node $i$, through a self-avoiding
walk after $h$ steps, is henceforth expressed as $P_h(i,j)$. In
order to estimate this probability, a total of $M$ self-avoiding
\emph{random} walks, starting from the node $i$ and proceeding $S$
steps, are performed. Note that these walks stop when one of the
following three conditions is met: (i) the walk reaches the maximum
pre-defined value of steps ($S+1$ nodes); (ii) the walk reaches an
extremity node, i.e., a node with degree one; or (iii) the walk
cannot proceed further because all of immediate neighbors of the
node at step $h$ were already visited. The probability $P_h(i,j)$
can then be estimated in terms of the number of times that the walks
departing from a node $i$ reach the node $j$ after $h$ steps,
divided by the number of walks, $M$. Note that $P_h(i,j)$ is
typically different of $P_h(j,i)$. After the probabilities are
estimated for each node, it is possible to calculate the
\emph{diversity entropy signature} $E_h(i)$ of a node $i$ after $h$
steps as~\cite{Costa2008a}:

\begin{equation}
E_h(i) = -\sum_{j=1}^{N}
\begin{cases}
           0         & \textrm{if $P_h(i,j) = 0$}\\
P_h(i,j)ln(P_h(i,j)) & \textrm{if $P_h(i,j) \neq 0$}
\end{cases}
\end{equation}

Although the diversity entropy provides an interesting quantification
of the accessibility of the nodes, the \emph{outward accessibility}
has been proposed as a normalization of the \emph{diversity entropy},
allowing direct comparison with other measures regarding non-linear
transient dynamics of the network (e.g. \emph{inward
accessibility}~\cite{Costa2008b}). The outward accessibility $OA_h(i)$
of a node $i$ after $h$ steps can be immediately calculated from the
diversity entropy as

\begin{equation}\label{eq:outward}
    OA_h(i) = \frac{exp(E_h(i))}{N-1}
\end{equation}

The outward accessibility was estimated for the central part of the
urban streets of S\~ao Carlos, which is represented by a network
with $N=2812$ nodes and $E=4713$ edges (Fig.~\ref{fig1}). The total
length of the self-avoiding random walks performed for each node was
$S=60$, and $10.000$ walks were performed so as to obtain accuracy
in probability estimations.  In Fig.~\ref{fig1} the gray levels of
the nodes corresponds to their respective accessibility, averaged
over all the steps. An interesting result which is evident in this
figure is that most part of the highly accessible nodes corresponds
to the downtown S\~ao Carlos, located at the central region of the
map. Another important property is the high spatial discriminative
power provided by the outward accessibility measurement: it can be
clearly seen from Fig.~\ref{fig1} that the nodes situated at the
border of the network have the smallest outward values, while the
inner nodes have the highest outward values. Interestingly, nodes
with low outward accessibility can be found even downtown.  It is
interesting to recall that the ability of the accessibility approach
to identify the most central (against the borders) parts of a
network is not restricted to geographical networks, but can be
immediately applied to any other type of complex network.

\begin{figure}[ht]
\includegraphics{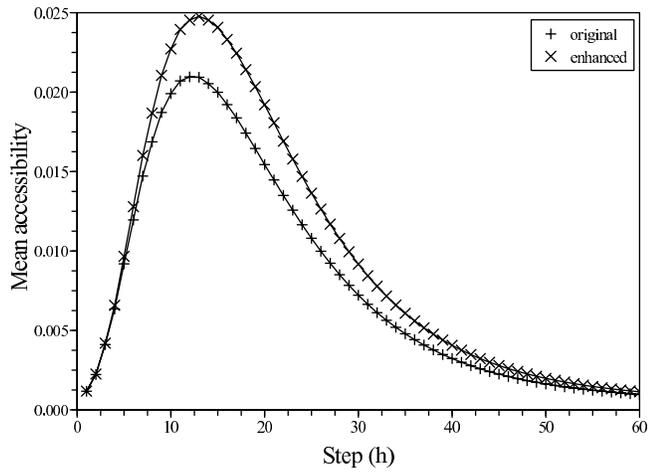}
\caption{Mean accessibility of the original network and the enhanced
network. In both cases, only the nodes with up to seven edges of
distance from the nodes that received the new connections were taken
into account. Note the substantial increase in the accessibility as
a consequence of the new connections.}\label{fig2}
\end{figure}

In order gain more insights regarding the accessibility of the
different parts of the town, we also considered hypothetical new
edges (i.e., new streets) connecting some nodes of the periphery and
internal regions of the town (represented as dashed lines in
Fig.~\ref{fig1}). This new arrangement allowed a study of the
potential impact of the new edges in the accessibility of their
neighborhood. The mean accessibility was computed considering the
nodes located up to seven blocks away from the nodes that received
the new connections. Figure~\ref{fig2} shows these values for the
original network and for the enhanced network.  Note an increase of
21\% in the accessibility after approximately $h=15$ steps for the
place where the new edges were added. This result shows that major
improvements of accessibility can be achieved by adding just a few
streets at strategic locations.

The authors thank the S\~ao Carlos town hall for providing and
granting the permission for using the city plan and Matheus P. Viana
for the design of the image processing routines. Bruno A. N.
Traven\c{c}olo is grateful to FAPESP for financial support
(2007/02938-5) and Luciano da Fontoura Costa thanks to CNPq
(301303/06-1) and FAPESP (05/00587-5) for financial support.

%\bibliography{inout}

\end{document}